\documentclass[%
preprint,
 amsmath,amssymb,
 aps,
showkeys,
notitlepage
]{revtex4-1}

\usepackage{etoolbox}

\makeatletter
\patchcmd{\frontmatter@abstract@produce}
  {\vskip200\p@\@plus1fil
   \penalty-200\relax
   \vskip-200\p@\@plus-1fil}
  {}
  {}
  {}
\makeatother

\usepackage{graphicx}
\usepackage{dcolumn}
\usepackage{bm}
\usepackage{hyperref}

\newcommand{\psr}{PSR~J1713+0747}

\begin{document}


\title{Testing fifth forces from the Galactic dark matter}
\thanks{Presented at the meeting {\it Recent Progress in Relativistic
Astrophysics}, 6-8 May 2019 (Shanghai, China).}%

\author{Lijing Shao}
 \email{lshao@pku.edu.cn}
\affiliation{%
 Kavli Institute for Astronomy and Astrophysics, Peking University,
Beijing 100871, China
}%

\date{\today}

\begin{abstract}
Is there an unknown long-range force between dark matter (DM) and
ordinary matters? When such a fifth force exists and in the case
that it is ignored, the equivalence principle (EP) is violated {\it
apparently}. The violation of EP was severely constrained by, for
examples, the E\"ot-Wash laboratory experiments, the lunar laser
ranging, the {\it MICROSCOPE} satellite, and the long-term
observation of binary pulsars. We discuss a recent bound that comes
from \psr{}. When it is combined with the other bounds, a
compelling limit on the hypothetical fifth force is derived. For
the neutral hydrogen, the strength of such a fifth force should not
exceed $1\%$ of the gravity.
\end{abstract}

\keywords{dark matter; equivalence principle; binary pulsars}
\maketitle


\section{Introduction}
\label{sec:intro}

Over the past hundreds of years, while physicists have established
a sophisticated picture to delineate the {\it ordinary} world around
us, we are still lacking a coherent description of the {\it dark}
world. Two notable substances, the dark matter (DM) and the dark
energy, were conjectured, though we do not know much detail of
them~\cite{Sahni:2004ai}. In this proceeding we focus on the DM. Up
to now, the DM was {\it solely} discovered via its gravitational
interaction with the ordinary matters. By using the word ``{\it
discovery}'', we mean to look for {\it interactions} with our
experimental instruments, either directly or indirectly. The
primary example for direct searches is to look for the interaction
of DM with nucleons in underground laboratories. As an example of
indirect searches, by looking for $\gamma$-ray excess in the
direction of the Galactic Center, we aim to detect DM particles
that, via some portal, decay or annihilate into some standard-model
particles which eventually couple to photons.
Although various means were performed for the past decades, and we have
learnt a lot from these direct and indirect experiments, no unanimously
accepted clues on the non-gravitational interaction were found yet, and the
nature of DM remains largely unknown~\cite{Bertone:2004pz}.

Most of past searches looked for possible short-range interactions
(say, via a massive force mediator) between the DM and the ordinary
matters. We here look for an alternative possibility. We
investigate the possibility that, besides the gravitational
interaction, there is an extra long-range force between the DM and
the ordinary matters~\cite{Stubbs:1993xk}. By saying
``long-range'', the force mediator should be massless or
ultralight, with its Compton wavelength $\lambda$ larger than the
typical length scale of the systems under
discussion~\cite{Adelberger:2009zz, Wagner:2012ui}. Here we make
use of the Galactic distribution of DM, hence $\lambda \gg {\cal
O}(10\,{\rm kpc})$ and the mass of the force mediator $m \ll
10^{-27} \, {\rm eV}/c^2$~\cite{Shao:2018klg}. 
We assume $m \to 0$ in the following study.

The spin-independent potential between body $A$ and the DM, from scalar
(``$-$'' sign) or vector (``$+$'' sign) exchange, is~\cite{Adelberger:2009zz,
Wagner:2012ui},
\begin{equation}
    V(r) = \mp g_5^2 \frac{q_5^{(A)} q_5^{({\rm DM})}}{4\pi r} \,,
\end{equation}
where $g_5$ is the coupling constant and $q_5$ is the dimensionless fifth-force
charge~\cite{Adelberger:2009zz, Wagner:2012ui}. If such a fifth force was
ignored by the experimenter, she/he will ``discover'' an {\it apparent}
violation of the equivalence principle (EP) between body $A$ and body $B$ when performing her/his gravity
experiments in the gravitational field of the DM, with an E\"otv\"os
parameter $\eta^{(A,B)}_{\rm DM}$~\cite{Stubbs:1993xk}, 
\begin{equation}
    \eta^{(A,B)}_{\rm DM} = \pm \frac{g_5^2}{4\pi G u^2}
    \frac{q_5^{\rm (DM)}}{\mu_{\rm DM}} \left[
    \frac{q_5^{(A)}}{\mu_A} - \frac{q_5^{(B)}}{\mu_B} \right] \,,
\end{equation}
where $G$ is the Newtonian gravitational constant, and
$\left( q_5 / \mu \right)$ is an object's charge per atomic
mass unit $u$. This is true even that the gravity is still
described by the general relativity (GR), and EP is valid if the
experimenter is aware of the fifth force. From observations, if
EP is observed to hold, one can put a limit on the fifth force.
Such tests were performed with the E\"ot-Wash laboratory
experiment~\cite{Stubbs:1993xk, Wagner:2012ui} and the lunar laser
ranging~\cite{Nordtvedt:1994ads, Williams:2009ads}. The E\"otv\"os
parameter was constrained to be $\left| \eta_{\rm DM} \right| \lesssim
10^{-5}$. Here we discuss an independent test from the binary
pulsar \psr{}~\cite{Shao:2018klg}, which has some specific
distinctions from Solar-system experiments.

The proceeding is organized as follows. In the next section the
relevant observational characteristics of \psr{} are
introduced~\cite{Zhu:2018etc}. In section~\ref{sec:ep}, we review
the EP-violating signal in the orbital dynamics of a binary
pulsar~\cite{Damour:1991rq, Freire:2012nb}. The method is applied
to put a limit on the fifth force in section~\ref{sec:5f} and the
advantages of using neutron stars (NSs) are outlined. The last
section discusses the possibility of finding a suitable binary
pulsar close to the Galactic Center, that will boost the test
significantly.

\begin{table}[b]
    \caption{Parameters for \psr{}~\cite{Zhu:2018etc}. The numbers in
    parentheses indicate the uncertainties on the last
    digit(s).\label{tab:psr}}
    \centering
    \begin{ruledtabular}
    \begin{tabular}{lr}
    \textbf{Parameter}	& \textbf{Value}	\\
    \hline
    Spin frequency, $\nu \, ({\rm s}^{-1})$ & 218.8118438547250(3) \\
    Orbital period, $P_b$\,(d) & 67.8251299228(5) \\
    Time derivative of $P_b$, $\dot P_b \, (10^{-12} \, {\rm s\,s}^{-1})$ & 0.34(15) \\
    Corrected $\dot P_b \, (10^{-12} \, {\rm s\,s}^{-1})$ & 0.03(15) \\
    Orbital inclination, $i$\,(deg) & 71.69(19) \\
    $\hat{\bm{x}}$ component of the eccentricity vector, $e_x$ & $-0.0000747752(7)$ \\
    $\hat{\bm{y}}$ component of the eccentricity vector, $e_y$ & $0.0000049721(19)$ \\
    Time derivative of $e_x$, $\dot e_x \, ({\rm s}^{-1})$ & $0.4(4) \times 10^{-17}$ \\
    Time derivative of $e_y$, $\dot e_y \, ({\rm s}^{-1})$ & $-1.7(4) \times 10^{-17}$ \\
    Companion mass, $m_c \, (M_\odot)$ & 0.290(11) \\
    Pulsar mass, $m_p \, (M_\odot)$ & 1.33(10) \\
    \end{tabular}
    \end{ruledtabular}
\end{table}

\section{\psr{}}
\label{sec:psr}

\psr{} is a 4.5\,ms pulsar in a binary system with an orbital
period $P_b = 68\,$d. Its companion is a white dwarf (WD) with mass
$m_c = 0.29\, M_\odot$. Due to its narrow pulse profile and stable
rotation, \psr{} is monitored by the North American Nanohertz
Gravitational Observatory (NANOGrav), the European Pulsar Timing
Array (EPTA), and the Parkes Pulsar Timing Array (PPTA).
\citet{Splaver:2004du}, \citet{Zhu:2015mdo}, and
\citet{Desvignes:2016yex} have published timing solutions for this
pulsar, and the latest timing parameters from combined datasets are
given in \citet{Zhu:2018etc}. Some relevant parameters for this
proceeding are collected in Table~\ref{tab:psr}.

Because of mass transfer activities in the past, this binary has a
nearly circular orbit. Nevertheless, its eccentricity, $e \lesssim
10^{-4}$, can still be measured. For the purposes in this study, we
define $e_x \equiv e \cos\omega$ and $e_y \equiv e\sin\omega$ where
$\omega$ is the longitude of periastron. Using data from 1993 to
2014, the timing precision of \psr{} has achieved to be sub-$\mu$s.
It renders a previous timing model for small-eccentricity binary
pulsars, {\sf ELL1}~\cite{Lange:2001rn}, not accurate enough.
\citet{Zhu:2018etc} developed an extended model, {\sf ELL1+}, by
including higher-order contributions from the eccentricity. The {\sf
ELL1+} model includes terms up to ${\cal O}\left( e^2 \right)$ in
the R\"omer delay~\cite{Zhu:2018etc}. The measured values for $e_x$
and $e_y$ are listed in Table~\ref{tab:psr}. In addition, the first
time derivatives of $e_x$ and $e_y$ are also given in the table,
assuming that the changes are linear in time~\cite{Zhu:2018etc}.

The measurements of the orbital decay and the eccentricity
evolution were used to put constraints on different aspects of
gravitational symmetries~\cite{Shao:2016ezh, Zhu:2018etc},
including
\begin{enumerate}
    \item the gravitational constant $G$'s constancy, $\left| \dot
    G / G \right| \lesssim 10^{-12} \, {\rm yr}^{-1}$;
    \item the universality of free fall for strongly
    self-gravitating bodies in the gravitational potential of the
    Milky Way, $\left| \eta_{\rm Gal} \right| < 0.002$; and
    \item the parameterized post-Newtonian (PPN) parameter $\hat
    \alpha_3$, $\left| \hat \alpha_3 \right| \lesssim 10^{-20}$.
\end{enumerate}
The first one is based on the orbital decay measurement, and the
rest are based on the eccentricity evolution
measurements~\cite{Zhu:2018etc}. The second test is our focus here
and it is to be discussed below.

\section{EP-violating signals in a binary pulsar}
\label{sec:ep}

\citet{Damour:1991rq} proposed to use small-eccentricity binary
pulsars in testing the strong EP. When the EP is violated, a
``gravitational Stark effect'' takes place and polarizes the binary
orbit in a characteristic way. 
A related phenomenon, the so-called ``Nordtvedt effect'', takes place
in the Earth-Moon-Sun system when the EP is violated~\cite{Nordtvedt:1968qs}.
It also happens for binary pulsars when the PPN preferred-frame parameters
$\alpha_1$~\cite{Damour:1992ah, Shao:2012eg} and
$\alpha_3$~\cite{Bell:1996ir, Zhu:2018etc} are nonzero. In the current case,
the relative acceleration between the pulsar and its companion star
reads~\cite{Damour:1991rq, Freire:2012nb},
\begin{equation}\label{eq:acc}
    \ddot{\bm{R}} = - \frac{ G M}{R^2} \hat{\bm{R}} + \bm{A}_{\rm
    PN} + \bm{A}_\eta \,,
\end{equation}
where $\bm{R}$ is the binary separation, $G$ is the gravitational
constant, $M$ is the total mass, and $\hat{\bm{R}} \equiv \bm{R}
/R$. The post-Newtonian (PN) corrections are collected in
$\bm{A}_{\rm PN}$, and the EP-violating abnormal acceleration is
denoted as $\bm{A}_\eta$. At leading order, $\bm{A}_\eta \simeq
\eta_{\rm DM} \bm{a}_{\rm DM}$ where $\bm{a}_{\rm DM}$ is the
gravitational acceleration generated by the DM. To be explicit,
here we take GR as the gravity, and the $\bm{A}_\eta$ term comes
from an unknown ``fifth force'' instead of some modified gravity.
If $\bm{A}_\eta$ comes from some modified gravity, then there will
be extra considerations; for example, in that case the gravitational
constant $G$ should be replaced with an effective gravitational
constant, ${\cal G}$~\cite{Freire:2012nb}.

We define the eccentricity vector $\bm{e}(t) \equiv e \hat{\bm{a}}$
to have a length of $e$, and a direction from the center of mass of
the binary towards the periastron, $\hat{\bm{a}}$. In the Newtonian
gravity, $\bm{e}(t)$ is a constant vector due to the fact that the
Newtonian interaction has a larger symmetry group than SO(3). In
GR, there is the famous periastron advance where, at leading order,
$\bm{e}(t)$ rotates uniformly at a rate,
\begin{equation}\label{eq:omdot:GR}
    \dot \omega_{\rm PN} = \frac{3}{1-e^2} \left( \frac{2\pi}{P_b}
    \right)^{5/3} \left( \frac{GM}{c^3} \right)^{2/3} \,.
\end{equation}
Under the relative acceleration (\ref{eq:acc}), equations of motion
get modified. After averaging over an orbit, the most important
ones read~\cite{Freire:2012nb},
\begin{align}
    \left\langle \frac{{\rm d} P_b}{{\rm d} t} \right\rangle =& 0 \,, \\
    \left\langle \frac{{\rm d} \bm{e}}{{\rm d} t} \right\rangle =& \bm{f} \times \bm{l} + \dot \omega_{\rm PN} \hat{\bm{k}} \times \bm{e} \,, \\
    \left\langle \frac{{\rm d} \bm{l}}{{\rm d} t} \right\rangle =& \bm{f} \times \bm{e} \,,
\end{align}
where $\hat{\bm{k}}$ is the direction of orbital norm, $\bm{l}
\equiv \sqrt{1-e^2} \hat{\bm{k}} $, and $\bm{f} \equiv 3
\bm{A}_\eta / \left( 16\pi G M / P_b \right)^{1/3}$. These
differential equations can be integrated to
give~\cite{Damour:1991rq, Freire:2012nb},
\begin{equation}\label{eq:et}
    \bm{e}(t) = \bm{e}_\eta + \bm{e}_{\rm GR}(t) \,,
\end{equation}
where $\bm{e}_{\rm GR}(t)$ is a uniformly rotating vector with a
rate according to GR's prediction (\ref{eq:omdot:GR}), and
$\bm{e}_\eta \equiv \bm{f}_\perp / \dot\omega_{\rm PN} $ is a
constant vector with $\bm{f}_\perp$ representing the projection of
$\bm{f}$ on the orbital plane.

\section{Constraints on the fifth force}
\label{sec:5f}

From the theoretical side, we have a characteristic evolution of
the eccentricity vector, dictated in Eq.~(\ref{eq:et}), while from
the observational side, we have measured the linear changes in the
eccentricity vector, decomposed to $\dot e_x$ and $\dot e_y$ in
Table~\ref{tab:psr}. Therefore, by comparing them, we can perform a
test of the existence of the $\bm{A}_\eta$ term. Notice that the DM
acceleration $\bm{a}_{\rm DM}$ comes from the Galactic DM
distribution. It is different from that of \citet{Zhu:2018etc}
where the authors, considering a different scenario, used the total
acceleration from the whole Milky Way to obtain $\eta_{\rm Gal}$.
We used an updated Galactic model~\cite{McMillan:2017} to calculate the
acceleration from the DM. This choice does not change the relative strength
in constraining the fifth force from different experiments.

The values of $\dot e_x$ and $\dot e_y$ from \psr{} are consistent
with $\bm{e}_\eta = 0$ in Eq.~(\ref{eq:et}), which means that
$\bm{A}_\eta = 0$ and $\eta_{\rm DM} = 0$. A careful analysis gives
$\left| \eta_{\rm DM} \right| < 4 \times10^{-3}$ at the 95\%
confidence level~\cite{Shao:2018klg}. This limit is weaker than
those from the E\"ot-Wash laboratory experiments and the lunar
laser ranging. However, due to the very nature of the celestial
binary system, \psr{} has multiple advantageous
aspects~\cite{Shao:2018klg}.
\begin{itemize}
    \item {\bf Driving force.} Because gravity is a manifest of the
    curved spacetime, free-fall states are ideal in performing
    gravity tests. Though the measurement precision is not as good
    as that of the E\"ot-Wash group, the {\it MICROSCOPE} satellite
    gains a factor of 500 in the driving force by putting the
    experiment in space in a free-fall state, thus achieving a
    better bound on $\eta$~\cite{Touboul:2017grn}. On the contrast,
    binary pulsars are usually worse in testing $\eta$ due to the
    smaller driving force from the Milky Way. Nevertheless, if
    considering the DM as the attracting source, binary pulsars do
    not have such a disadvantage, for all the experiments performed
    in the Solar system have the same attraction from the Galactic
    DM distribution. Even better, if a suitable binary pulsar is
    found in a region that has a larger DM attraction, it may
    outperform the other tests (see the next section). It is interesting to note that, the triple pulsar, PSR~J0337+1715, though being excellent in testing the strong EP~\cite{Archibald:2018oxs, Shao:2016ubu}, does not probe the fifth force from the Galactic DM.
    \item {\bf Measurement precision.} The precision in measuring
    $\dot{\bm{e}}$ is proportional to $\sigma / \sqrt{\bar N T^3 }$
    where $\sigma$ is the rms of time-of-arrival (TOA) residuals,
    $\bar N$ is the average number of TOAs per unit time, and $T$
    is the observational baseline~\cite{Freire:2012nb}. Therefore,
    the test from binary pulsars will improve as a function of
    time, especially with the new instruments, like the
    Five-hundred-meter Aperture Spherical radio Telescope (FAST) in
    China~\cite{Nan:2011um}, and the Square Kilometre Array (SKA)
    in Australia and South Africa~\cite{Kramer:2004hd,
    Shao:2014wja, Bull:2018lat}.
    \item {\bf Material sensitivity.} Unlike the majority of solid
    materials on the Earth that have similar portions of protons
    and neutrons, NSs are almost 100\% made of neutrons which are
    different from its WD companion. This gains a factor of ${\cal
    O} \left( 10^2 \right)$ when interpreting the result $\left|
    \eta_{\rm DM} \right| < 4 \times10^{-3}$ to more fundamental
    theory quantities. Thus, though the measurement of $\eta_{\rm
    DM}$ from \psr{} is worse than the other measurements, it has a
    comparable power when being translated into fundamental theory
    parameters (see Fig.~1 in Ref.~\cite{Shao:2018klg} for
    details).
    \item {\bf Binding energy.} Ordinary materials that were used
    in the EP test have a mass deficit about ${\cal O}
    \left(0.1\%\right)$ due to the nuclear binding energy. NSs,
    being strongly self-gravitating, have a mass deficit about
    ${\cal O} \left(10\%\right)$ due to the gravitational binding.
    This benefits a lot in probing some specific parameter space
    that is very hard to investigate with solely terrestrial
    experiments (see Table~1 and Fig.~1 in
    Ref.~\cite{Shao:2018klg}).
\end{itemize}

By combining all existing EP experiments, we reach the following
conclusion: {\it if there is a long-range fifth force between the
DM and the ordinary matters, its strength should not exceed $1\%$
of the gravitational force for neutral
hydrogens}~\cite{Shao:2018klg}.

\section{Galactic Center binary pulsars}
\label{sec:gc}

As is discussed in the previous section, the driving force from the
DM is important in this test. The experiments in the Solar system,
by definition, cannot be done elsewhere but in the Solar system.
Due to the static large-scale DM distribution in the Galaxy, the
acceleration $\bm{a}_{\rm DM}$ in the Solar system is a fixed
quantity, almost zero variation from place to place inside the
Solar system. Therefore, for these experiments, one cannot enlarge
its driving force.

However, binary pulsars in principle can be distributed anywhere in
the Galaxy, and in the future that the SKA is to discover {\it all}
pulsars in the Milky Way that point towards the
Earth~\cite{Kramer:2004hd}. Among them, it is likely that there are
suitable binary pulsars for this test in the region where the
driving force is much larger. In particular, we consider the
Galactic Center region where the DM density is much denser.
\citet{Gondolo:1999ef} argued that in the inner region of our
Galaxy, there might be a DM spike. Such a spike will indeed enhance
the driving force significantly when a binary pulsar has a distance
smaller than $\sim 10\,$parsec to the Galactic Center. Studies on
the pulsar population suggested that the inner parsec could harbor as
many as thousands of active radio pulsars that beam at the
Earth~\cite{Wharton:2011dv}. Current and future searching plans
are on their way (see e.g. Ref.~\cite{Goddi:2017pfy}).

\begin{acknowledgments}
We thank Zhoujian Cao, Michael Kramer, and Norbert
Wex for helpful discussions.
This work was supported by the Young Elite Scientists Sponsorship
Program by the China Association for Science and Technology
(2018QNRC001), and partially supported by the National Natural
Science Foundation of China (11721303), and the Strategic Priority
Research Program of the Chinese Academy of Sciences through the
Grant No. XDB23010200.
\end{acknowledgments}

%

\end{document}